\documentclass[a4paper,12pt,epsfig,superscriptaddress,showpacs]{article}
\usepackage{cite} 
\usepackage[utf8]{inputenc}
\usepackage{graphicx}
\usepackage{amsmath}
\usepackage{amssymb}
\usepackage{cmap}
\usepackage{bm}
\usepackage{hyperref}
\usepackage{array}
\usepackage{epsfig}
\usepackage{array}
\usepackage{color}
\usepackage{cite}
\usepackage{latexsym}
\usepackage{amscd, xypic}

\usepackage[left=1.0in, right=1.0in, top=1.0in, bottom=1.0in]{geometry}

\title{\bf Nonperturbative SU(3) thermodynamics  and the phase transition}

\author{
N.O.~Agasian$^{a,b,}$\thanks{agasian@itep.ru}\,\,, M.S.~Lukashov$^{a,c,}$\thanks{lukashov@phystech.edu}\,\, and Yu.A.~Simonov$^{a,}$\thanks{simonov@itep.ru}\, \\
\\
$^a$ \small{\em Alikhanov Institute for Theoretical and Experimental Physics,}\\
\small{\em Moscow 117218, Russia}\\
$^b$ \small{\em National Research Nuclear University ``MEPhI'',}\\
\small{\em Moscow 115409, Russia}\\
$^c$ \small{\em Moscow Institute of Physics and Technology,}\\
\small{\em Dolgoprudny 141700, Moscow Region, Russia}
}\bigskip
\date{\today}

\newcommand{\be}{\begin{equation}}
\newcommand{\ee}{\end{equation}}

\def\la{\mathrel{\mathpalette\fun <}}
\def\ga{\mathrel{\mathpalette\fun >}}
\def\fun#1#2{\lower3.6pt\vbox{\baselineskip0pt\lineskip.9pt
\ialign{$\mathsurround=0pt#1\hfil ##\hfil$\crcr#2\crcr\sim\crcr}}}

\newcommand{{\SD}}{\rm SD}

\newcommand{\vex}{\mbox{\boldmath${\rm x}$}}

\newcommand{\ver}{\mbox{\boldmath${\rm r}$}}

\newcommand{\veP}{\mbox{\boldmath${\rm P}$}}
\newcommand{\vep}{\mbox{\boldmath${\rm p}$}}

\newcommand{\vez}{\mbox{\boldmath${\rm z}$}}

\newcommand{\veu}{\mbox{\boldmath${\rm u}$}}

\newcommand{{\Mc}}{\mathcal{M}}

\newcommand{\lan}{\langle}
\newcommand{\ran}{\rangle}

\begin{document}

\maketitle
\begin{abstract}
The $SU(3)$  equation of state  ($P(T),\,s(T),\,I(T)$) are calculated within the
Field Correlator Method both in the confined and the deconfined phases. The
basic dynamics in our approach is contained  in the vacuum  correlators,  both
of the colorelectric (CE) and colormagnetic (CM) types, which ensure CE and CM
confinement below $T_c$ and CM confinement and Polyakov loops above $T_c$. The
resulting values of $T_c$
 and $P(T),\,I(T)$, $s(T)$ are in good  agreement with lattice
measurements.
\end{abstract}

\newpage

\section{Introduction }

The dynamics  of QCD at  small temperatures is known to be governed by
confinement, which establishes its scale, connected to the string tension
$\sigma$, and  this scale defines the nucleon mass and the  most energy density
of the visible part of the Universe.

The theory of confinement based on the vacuum averages of  field correlators in
QCD, was suggested in  \cite{12}, see \cite{13}  for reviews.

The idea that QCD might have a different phase without confinement  at large
temperature, was suggested long ago \cite{3*,4*}.

This deconfinement phase was studied  in the same framework of the vacuum
correlators, soon  after the theory of confinement in  \cite{8}, and it was
finally elaborated in \cite{9,10, 11*}, see the review in \cite{11}, where
numerical calculations were done and compared to  existing data. The theory of
temperature transition in QCD given in \cite{9,10} is easily generalized to the
case of nonzero density \cite{11*}.

The main idea of the temperature transition in QCD given in all these papers is
based on two points:

1) From the  basic thermodynamics law one can deduce, that the  states with the
minimal free energy (maximal pressure)  are more probable. Therefore with the
  growing temperature the physical systems prefer    configurations with
  reduced  correlators and larger entropy.  As  the consequence the phase with zero colorelectric
confining vacuum correlators and condensates (and nonzero colormagnetic)  wins
at some temperature, leading to the deconfining vacuum.

2) The lowest (also the dominant) Gaussian field correlators provide two basic
interactions: the linear  confining $V_D^{\rm lin}(r)  \sim \sigma r$ and two
interactions with saturating maxima: $V_1 (r, T)$  and   $V_D^{\rm sat} (r,T)$
  where $V (\infty, T) $  const.  The latter   yields
automatically the Polyakov lines $L_a (T) = \exp \left( -\frac{c_a V_1 (\infty,
T)}{2T} \right), ~ c_3 =1, ~ c_8 =\frac94$, which enter linearly the
thermodynamic potential and suppress its magnitude. This is a basic point,
since in our approach $L_a (T)$ appear necessarily in $F(T)$ as factors  in the
deconfinement phase, and  it is not a model assumption.

As it  was shown in \cite{10}, $L_a (T)$ alone give  a  reasonable (within
20-25\%) description of the $P(T), I(T)$ etc. in the  deconfined phase,  when
all other nonperturbative (e.g. colormagnetic)  contributions are neglected.

In addition, this  lowest approximation used in \cite{10},  with free gluon and
quark loops augmented by known Polyakov loops   was able to predict the main
rough characteristics, transition (crossover) temperature $T_c$ and even its
chemical potential dependence $T_c (\mu)$ \cite{11*}, as well  as pressure
$P(T)$, trace anomaly $I(T) =\varepsilon - 3 P$,  sound velocity $c_s(T)$
\cite{10} etc. with reasonable accuracy.

An interesting development of the same
 deconfinement theory is contained in \cite{12*,13*}, where the influence of
 strong magnetic fields was taken into account, again in good  agreement with
 lattice  data.

It is a purpose of the present paper to make a    step further, and to take
into account another important  nonperturbative (np) interaction: the
colormagnetic confinement with the string tension $\sigma_s$. It was shown in
\cite{25} that it resolves the Linde problem \cite{23,24} and  creates bound
states in 3d \cite{34}. Here we would like to study how it affects the pure
SU(3) thermodynamic potentials, in particular $P(T)$,  $I(T)$, latent heat,
critical temperature $T_c$.

One of advantages of our analytic approach is that we can analyze the $N_c$
behavior   of all quantities and compare it to numerical studies \cite{19*,
16*}.

 The SU(3) gluodynamics is an important testing ground for the theory, since it
 contains most np and perturbative characteristics of the full
 QCD. On the    lattice side already the  first studies
 \cite{1,2,3} revealed the phase transition and important new physical effects
 both   below and above $T_c$. On the perturbative side the resummation method
 of  the Hard Thermal Loop (HTL), first developed in \cite{4,5}, was used in
 \cite{6,7} in the SU(3) theory, demonstrating a good agreement with lattice
 data
 at large $T$, whereas at $T< 4 T_c $ one needs $np$ contributions. On the
 lattice side the most accurate data are obtained in \cite{43}, see also
 \cite{24a} for a recent publication. In an alternative way the $SU(3)$ thermodynamics was studied in the framework  of effective theories in \cite{Yaffe:1982qf, Vuorinen:2006nz, Ogilvie:2012is, Meisinger:2001yk, Andreev:2007zv, Ratti:2006wg, Ratti:2005jh, Rossner:2007ik}, in particular in the PNJL model in \cite{Ratti:2006wg, Ratti:2005jh, Rossner:2007ik}, while in \cite{Andreev:2007zv} the author exploited the AdS/QCD formalism.


 In  what follows we shall start from  the theory  developed in  \cite{8,9,10}, but make more explicit
 the dynamics in the  confined and deconfined phases.

 Note, that the basic ground for this deconfinement theory is  already contained in the
 np confinement mechanism, suggested in \cite{12}.

 In this approach the confinement is a result of the np color field
 correlators, which are vacuum averages of the Euclidean colorelectric (CE) and
 colormagnetic (CM) field  $\lan tr E_i (x) E_J (y)\ran $, ~$\lan tr H_i (x)
 H_j(y) \ran$,    ~proportional ~ to ~ functions ~(correlators)    ~$D^E (x-y),$\\$ D_1^E
 (x-y)$   and $D^H (x-y),$ $ D_1^H(x-y)$ respectively.
 $$ \frac{g^2}{N_c}\lan\lan {\rm
Tr} E_i(x)\Phi
E_j(y)\Phi^\dagger\ran\ran=\delta_{ij}\left(D^E(u)+D_1^E(u)+u^2_4\frac{\partial
D_1^E}{\partial u^2}\right)+ u_iu_j\frac{\partial D_1^E}{\partial u^2},$$ \be
\frac{g^2}{N_c}\lan\lan {\rm Tr} H_i(x)\Phi
H_j(y)\Phi^\dagger\ran\ran=\delta_{ij}\left(D^H(u)+D_1^H(u)+\veu^2\frac{\partial
D_1^H}{\partial\veu^2}\right)- u_iu_j\frac{\partial D_1^H}{\partial
u^2},\label{0a}\ee

Here $ u=x-y$  and $\Phi(x,y) = P\exp (ig \int^x_y A_\mu dz_\mu)$ is the
parallel transporter, needed to maintain the gauge invariance of  relations
(\ref{0a}).

 The confining correlators $D^E, D^H$ generate  the nonzero values of CE and CM string tensions,
 \be \sigma^{E(H)} = \frac12 \int D^{E (H)} (z) d^2z.\label{1a}\ee
 At zero temperature $T$ both string tensions coincide and $\sigma^E$ forms  the basic np
 scale, which defines all hadron  masses and the  QCD scale in general.

 To make the theory selfconsistent, one must calculate $D^{E(H)}, D_1^{E(H)},$
 via $\sigma^E=\sigma^H \equiv \sigma$  and prove that Eq.(\ref{1a}) is
 satisfied. This was done in \cite{15}, where it was shown that the correlators
 are  proportional to  the Green's  functions of gluelumps, calculated before on the
 lattice \cite{16} and analytically in the framework of our method \cite{17}.

The correlators $D^E$ and $D^E_1$ produce both the scalar confining interaction
$V_D(r)$ and the vector-like interaction $V_1(r)$.

\be V_D(r) = 2 c_a \int^r_0 (r-\lambda) d\lambda \int^\infty_0 d\nu D^E
(\lambda, \nu)= V_D^{(\rm lin)} (r)+ V^{(\rm sat)}_D (r) \label{3b}\ee
 \be V_1 (r)= c_a
\int^r_0 \lambda d\lambda \int^\infty_0 d\nu D_1^E (\lambda, \nu), ~~c_{\rm
fund} =1, ~~ c_{\rm adj} =9/4.\label{4b}\ee

Separating from $V_D(r)$ the purely linear form $V_ D^{(\rm lin)} (r)$ and
using the renormalization procedure for $V_1(r)$  with account of the
perturbative gluon exchange, $V_1 (r) =V^{\rm sat}_{1} (r)+ V_{OGE} (r)$, one
obtains the general structure of the $q\bar q$ or $gg$ interaction in the
region $T<T_c$.

\be V(r, T<T_c)= V^{\rm lin}_D (r) + V_D^{\rm sat} (r) + V_1^{\rm sat} (r) +
V_{OGE} (r). \label{5b}\ee

It is interesting, that both parts, $V_D^{\rm sat} + V^{\rm sat}_{1}$,
saturating at large $r$,   compensate each other at small $T$, as shown in
appendix, and one is retained with the standard linear + OGE interaction, in
exact agreement with lattice and experiment.

However at $T\geq T_c$, when $D^E$ vanishes, one obtains two terms, $V_1^{\rm
sat} $ and $V_{OGE}$, which together with $\sigma_s$ define the dynamics.

 The np thermodynamics  \cite{10,11} based on the  field correlators (FC),
 considers the low temperature phase of SU(3), and of QCD in general, as the
 confined phase, where thermal degrees of freedom are white hadrons, glueballs
 in the SU(3) case, where all FC $(D^E, D^E_1, D^H, D^H_1)$ are nonzero and
 therefore both CE and CM (spatial) confinement are present.

Since $D^E_1$ is nonzero  above $T_c$, one may associate with it and  with
$D^H, D^H_1$  the deconfined phase (phase II), while the confined phase (phase
I) contains all four correlators $D^E, D^E_1, D^H, D^H_1$, so that the phase
transition  can be found from the intersection of two curves $P_I(T)$ and
$P_{II}(T)$, as  shown in Fig.1 and will be demonstrated below.

  In phase I the special role is  played by $D^E(\sigma^E)$, which  ensure not only
 confinement in the  usual sense, but also chiral symmetry breaking (CSB),
 \cite{30*}. As mentioned above,  the nonzero np part of $D^E_1$ is almost
 totally  compensated by $D^E$ for $T<T_c$, while the perturbative part yields
 gluon exchange contribution.
  The CM correlators $D^H, D^H_1$ ensure most part
 of
 spin-dependent forces \cite{31*} and CM confinement.

 With  the growth of $T$  for $T<T_c$ nothing special happens, except that more
 and more excited states (glueballs in SU(3)) participate  in the partition
 function, ensuring a steady but slow increase of  the pressure $P_{\rm conf}\equiv P_I(T)$ with
 $T$. This corresponds to the vacuum with all correlators nonzero.

 An interesting feature of the  glueball pressure $P_{\rm conf}$ is that the
 standard Hardron Resonance Gas (HRG) approach is not able to  sustain  the
 growth of $P_{\rm conf}$ near $T_c$ and one is using the Hagedorn
  enhancement in addition  to HRG  to comply with the  lattice
 data. We show in the  paper, that instead of the Hagedorn  factors,
 which  we consider   inappropriate to us, as will be discussed below, one can use the effect of string tension
 damping with  temperature near $T_c$, observed on the lattice \cite{29*,29**,
29***}, which strongly increases $P_{\rm conf}$  at $T\la T_c$and brings it in
agreement
 with  lattice data \cite{43}.

   The deconfined phase (phase II)   corresponds   to the zero values of
   $D^E$ and $\sigma^E$, and nonzero $D_1^E, D^H, D_1^H$. In this case the
   physical degrees of freedom are gluons, interacting via these correlators.
   At $T=T_c$ the fast growing $P_{\rm dec}$ keeps up with $P_{\rm conf}$ and the phase
   transition occurs,as it is shown in Fig. \ref{fig:fig01}.

{
\begin{figure}[htb] 
\setlength{\unitlength}{1.0cm}
\centering
\begin{picture}(6.8,6.8)
\put(0.6,0.5){\includegraphics[height=5.7cm]{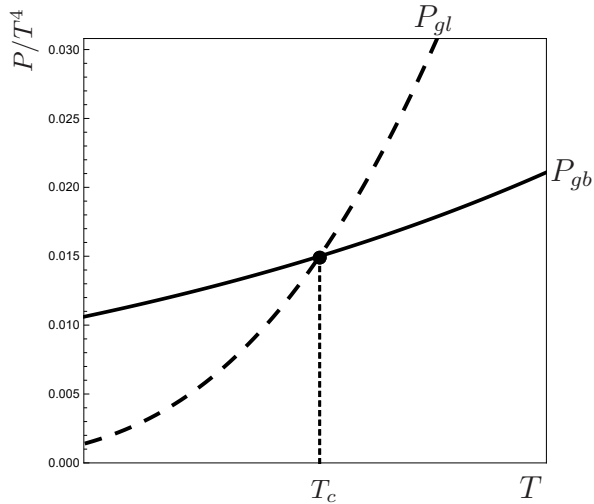}}
\put(6.7925,0.1){$T$}
\put(4.05,0.1){\footnotesize $T_c$}
\put(0.1,5.75){\footnotesize \rotatebox{90}{$P/T^4$}}
\put(5.40,6.35){$P_{gl}$}
\put(7.20,4.3){$P_{gb}$}
\end{picture}
\caption{Pressure $P(T)$ as function of temperature $T$ for the confined phase (glueballs) -- solid line, and for the deconfined phase (dashed line). The intersection point is at the critical temperature $T_{c}$.}
\label{fig:fig01}
\end{figure}
}\medskip


   One should stress the important role of $ V^{\rm sat}_{1}$, which  is compensated  by $ V^{\rm sat}_{D}$
   at $T<T_c$ (see appendix), but creates  its own pair interaction $V_1(r,T)$ for $T>T_c$
   \cite{10,18}, with nonzero value at $r\to \infty, ~~ V_1 (\infty, T)$. This
   term produces  the Polyakov loop of gluon $L_{\rm adj} (T) =
   \exp\left( - 9 \frac{ V_1 (\infty, T)}{8T}\right),$ and $ L_{\rm adj} = (L_f)^{9/4} $
    increases with $T$ and tends to constant for $T\la 2T_c$. This picture was
    successfully  confronted with lattice data in \cite{18}.

    One should note at this point, that $L_{\rm adj} (T)\equiv L_{\rm adj} $ remains nonzero  in the  confined phase for
    $T<T_c$, where it  is expressed via the  gluelump mass $m_{glp}\approx 1$
    GeV, $L_{\rm adj}^{<} (T) \cong \exp \left(-\frac{m_{glp}}{T}\right)$, and thus
    $L_{\rm adj} (T), T<T_c$ is much smaller than $L_{\rm adj} (T>T_c)$, in agreement
    with lattice data \cite{19}, as it was shown in the second refs. in \cite{10}.

However this  $L_{\rm adj}^{<} (T)$ does not enter the thermodynamic potential
of the confined phase and its properties are not of  interest for us.

    This general picture of the temperature dependence of FC and $\sigma^E,
    \sigma^H$ is  in agreement with  lattice measurements of the  correlators
    in \cite{20}, which demonstrate, that only correlator $D^E$ vanishes at $T\geq T_c$.

Till now nothing was said about the role of the spatial string tension
    $\sigma_s \equiv \sigma^H$ and the magnetic confinement in general in the  deconfinement transition. In the
    confinement
    region $T<T_c$, magnetic confinement is acting mostly in the  hadrons with
    angular momentum $L>0$, where it gives a small correction \cite{21}. In the
     deconfined region the situation is different. Here closed loop trajectories
     of gluons and quarks  for large
     $T$ lie almost all in $d=3$ space, and  therefore governed by the  spatial
     confinement growing with $T$. This provides every gluon with an  effective
     mass $m_{gl}$ proportional to $\sqrt{\sigma_s (T)}$.


     The same happens with space-like gluons, exchanged by the quark or gluon
     currents, those acquire the np Debye mass $m^H_{D} \approx 2
     \sqrt{\sigma_s(T)}$ \cite{22, 24*}. This phenomenon lifts  the IR divergences in the perturbative thermal series, noted in
     the well-known Linde
     problem \cite{23,24}, as it is explained in a recent paper \cite{25}, see
     also \cite{9} for an earlier discussion. At this point one should stress,
     that as found from $d=3$ SU(3) and on the lattice \cite{26}, also
      within our method as shown  in \cite{25}, $\sigma_s(T)$ is growing with $T$ as  $\sigma_s (T)
     = c^2_\sigma g^4 (T) T^2$, and hence in our np method  the CM  gluon screening  masses scale as
     $m_{gl} \sim g^2 (T) T$, whereas in the perturbative theory  the
     effective gluon mass  is of the CE origin $m_D^E (T) \sim g T + O(g^2)$,where $O(g^2)$ is of the np
     origin.

     From the practical point of view both definitions of the effective gluon
     mass are close numerically, since $g(T) \sim O(1)$ for $T \sim (300-500) $
     MeV, and therefore an average gluon mass, entering in HTL \cite{5,6,7} approach, may be
     not far from the magnetic $m^H_D$ \cite{22}.

     It is a purpose of the present paper to study the SU(3) thermodynamics in
     the lowest np approximation (the so-called Single-Loop Approach (SLA)) but
     taking into account the np correlators $D^E_1$ and $D^H$ for $T>T_c$,
     which produce Polyakov loops and $\sigma_s$ respectively. We calculate
     from $\sigma_s$ the gluon effective mass and  find $P(T), I(T) =
     \varepsilon-3P$.  We define
     $T_c$, latent heat and other characteristics and compare our results  to the  recent lattice  measurements in
        \cite{43}.

     The paper is organized as follows. In the next section the general field
     correlator formalism for thermodynamics is shortly summarized. In section
     3 the effect of magnetic confinement contributions is studied and
     estimated in the SLA approximation.
     Section 4 comprises the notion and numerical estimates of Polyakov loops, in comparison with lattice data.
      Section 5 is
     devoted to the discussion of the  confinement  phase   and  the  temperature dependence of the glueball pressure,
      in section 6 the results of the calculation of $T_c$,  pressure,
     and trace anomaly   are given,
     while  the  section 7 contains  a summary and prospectives.

     \section{General formalism}

     We are using the thermal background perturbation theory for the gluons in the  deconfined phase II,
     developed in \cite{10}, where vacuum background fields are denoted by
     $B_\mu$ and perturbative part by $a_\mu$. To the lowest order in $ga_\mu$
     one can write for the $B$ dependent free energy

     \begin{eqnarray}
\frac{1}{T} F_0^{gl}(B) &=&\frac12\ln \det G^{-1} -\ln \det (- D^2(B))=
\nonumber
\\
&=&Sp\left\{ -\frac12 \int^\infty_0 \xi (s) \frac{ds}{s} e^{-sG^{-1}} +
\int^\infty_0 \xi (s) \frac{ds}{s} e^{sD^2(B)}\right\}, \label{1}
\end{eqnarray}
while the vacuum averaged free energy is \be -\frac{\lan F_0^{ gl}
(B)\ran_B}{T}=\ln \left\lan \exp \left(-\frac{\lan F_0^{ gl}
(B)\ran}{T}\right)\right\ran_B.\label{2}\ee

Using the cluster expansion  in the exponent
\begin{eqnarray}
\lan \exp f\ran_B&&=\exp \left (\sum^\infty_{n=1} \lan \lan
f^n\ran\ran\frac{1}{n!}\right) \nonumber
\\
&&=\exp \{ \lan f\ran_B+\frac12[ \lan f^2\ran_B- \lan f\ran^2_B]+O (f^3)\},
\label{3}
\end{eqnarray}
one obtains the lowest order one-loop expression for $\lan F_0^{gl} (B)
\ran_B$,

\be
 \lan F_0^{gl}(B) \ran_B= -T\int\frac{ds}{s} \xi(s) d^4
 x(Dz)^w_{xx} e^{-K}\left [ \frac12 tr \lan \tilde
 \Phi_F(x,x)\ran_B-\lan tr \tilde \Phi (x,x)\ran_B\right].
 \label{4}
 \ee
 Here the winding path integration is

 \begin{eqnarray}
(Dz)^w_{xy}&&= \lim_{N\to \infty}\prod^N_{m=1}
\frac{d^4\zeta(m)}{(4\pi\varepsilon)^2} \nonumber
\\
&& \sum_{n=0,\pm,...} \frac{d^4p}{(2\pi)^4}\exp \left[ ip_\mu\left(
\sum^N_{m=1} \zeta_\mu(m)-(x-y)_\mu-n\beta \delta_{\mu 4}\right)\right].
\label{5}
\end{eqnarray}
and $\tilde \Phi(x,x) $ is the adjoint parallel transporter  \be \tilde
\Phi(x,y) = P\exp (ig\int^x_y \tilde B_\mu dz_\mu),\label{6}\ee while $\tilde
\Phi_F$ contains additional gluon spin factor, $P_F \exp (2ig \int^s_0 \tilde F
d\tau),$ which we shall replace by unity in the lowest approximation\footnote{
Here $P,P_F$ are ordering operators for the fields $\tilde B_\mu$ and $\tilde
F_{\mu\nu}$ respectively}. As a result the gluon pressure $P_{gl} V_3 = - \lan
F_0^{gl} (B) \ran_B$ can be written as \be P_{gl} = (N_c^2-1) \int^\infty_0
\frac{ds}{s} \sum_{n=0,\pm 1,\pm 2,...} G^{(n)} (s).\label{7}\ee

 $G^{(n)}$ in (\ref{7}) is defined as \be G^{(n)} (s) = \int (Dz)^w_{on}
e^{-K} \lan \hat{tr}_a W (C_n)\ran,\label{8}\ee where \be K=\frac14 \int^s_0
\left( \frac{dz_\mu (\tau)}{d\tau} \right)^2 d\tau,\label{9}\ee \be \lan
\hat{tr}_a W(C_n)\ran = \frac{tr_a}{(N_c^2-1)} \lan \tilde \Phi
(x,x^{(n)})\ran.\label{10}\ee

Note here, that the generic path of the gluon starts at the point $x$ and  ends
at the point $x^{(n)} = x_\mu+ n \beta\cdot \delta_{\mu 4}$, as shown in
(\ref{5}), so that one has a closed loop in 3d, while the projection on the
4-th axis yields the Polyakov loop, $L_{\rm adj}$. Indeed, for the propagator
$G(x,y)$ the Matsubara assignment in (\ref{5}) yields a sum of end points
$y^{(n)}_4 =y_4 +n\beta, n=0,\pm  1,...$ which  for the coinciding $x_4=y_4$
results in  an infinitive series of open contours $[y_4, y_4+n\beta]$, with the
unitary gauge equivalent points $U(y_4 +n\beta) = U(y_4)$. Now multiplying the
contours with  the product of gauge invariant lines  (\ref{6}), $\tilde \Phi
(y_4, y_4+n\beta) \times \tilde \Phi (y_4+n\beta,y_4  )=1$, and taking the
vacuum average,  one obtains the product of the closed Wilson loop $W_3$ and
the Polyakov line $L_{adj} (T)$ (modulo insignificant correlation between the
CE contents of $L_{adj}$ and CM of $W_3$).

As a result (\ref{10}) can be written  as \be \frac{tr_a}{(N_c^2-1)} \lan
\tilde \Phi (x,x^{(n)})\ran= L_{\rm adj}^{(n)} (T) \lan W_3\ran,\label{11}\ee
where $\lan W_3\ran$ is the spatial area law  factor \be \lan W_3 \ran=\exp
(-\sigma_s A_3)\label{12}\ee and $A_3$ is the minimal area in the 3d space of
the loop, formed by trajectories $z_i (\tau), 0\leq\tau \leq s, i=1,2,3.$

It will be essential that $\sigma_s(T)$ grows with $T$ as \cite{26, 25}
  \be \sigma_s (T) =
c^2_\sigma g^4 (T) T^2, \label{13}\ee where $c_\sigma$ is a dimensionless
constant defined in a np way. The form (\ref{13}) was found on the lattice
\cite{26} with $c_\sigma = 0.566\pm 0.013$. The similar form was found in $d=4$
\cite{24*,25}, using the gluelump Green's function method \cite{16,17}. For $T<
T_c, \sigma_s$ tends to a constant $\sigma_s = \sigma^{(E)}.$

We turn now to  the first factor on the r.h.s. of (\ref{11}). As it is shown in
\cite{ 10},    for $T>T_c$ one can express $L_{\rm adj}^{(n)}$ via the CE
correlator $D_1^E(z)$, \be L_{\rm adj}^{(n)} =\exp \left( - \frac{9}{4}
J^E_n\right), ~~ J_n^E = \frac{n\beta}{2} \int^{n\beta}_{0} d\nu\left(
1-\frac{\nu}{n\beta}\right) \int^\infty_0 \xi d\xi D_1^E
(\sqrt{\xi^2+\nu^2})\label{14}\ee  It is argued in \cite{10}, that a good
approximation for $T<1 $ GeV is $ J_n^E \cong n J_1^E$, which we shall use in
what follows.

The integral $(Dz_4)^w_{on} $ in (\ref{8}) for $T>T_c$ can be done  explicitly,
yielding \cite{10} \be G^{(n)}(s) = \frac{1}{\sqrt{4\pi s}}
e^{-\frac{n^2}{4T^2s} } G_3  (s) L_{\rm adj}^{(n)},\label{15}\ee where
$G_3(s) $ is \be G_3(s)= \int (D^3 z)_{xx} e^{-K_{3d}} \lan
W_3\ran,\label{16}\ee and as a result the gluon pressure in the phase II has
the form \be P_{gl} = \frac{N_c^2-1}{\sqrt{4\pi}} \int^\infty_0
\frac{ds}{s^{3/2}} G_3 (s) \sum_{n=0, 1,2,..} e^{-\frac{n^2}{4T^2s} } L_{\rm
adj}^{(n)}.\label{17}\ee

 \be  F_{gl}  = - P_{gl}V_3, \label{18}\ee

\section{Calculation of the spatial loop}

We consider here $G_3 (s)$, Eq.(\ref{16}), which corresponds to  the 3d loop,
which is governed by the spatial confinement  with the string tension
$\sigma_s(T)$. It is clear, that gluons on the opposite sides of the loop are
connected by the confining string,  and  we transform the integral (\ref{16})
to  make it explicit. To this end  we write the identity \be (D^3z)_{xx} =
(D^3z)_{xu} d^3u(D^3z)_{ux},\label{24}\ee where we choose the point $u_i$ as
$u_i = z_i \left( \frac{s}{2}\right).$

{
}

Using $u_3\equiv t$ as the Euclidean time in 3d, one can write \be (Dz_3
)_{x_3u_3} e^{-K_3} = \frac{1}{\sqrt{2\pi s}}, ~~ K_3 = \frac14 \int^{s/2}_0
\left( \frac{dz_3}{d\tau} \right)^2 d\tau.\label{25}\ee

As a result $G_3  (s)$ acquires the form \be G_3  (s) = \int (D^2z)_{xu} d^2u (
D^2z)_{ux} e^{-K_1-K_2} \lan W_3\ran \frac{dt}{2\pi s}.\label{26}\ee

Using (\ref{12}) one can express $\lan W_3\ran $ in terms of the instantaneous
confining potential $V_{\rm conf} = \sigma_s | \ver_1 -\ver_2|, \lan W_3\ran =
\exp (- V_{\rm conf} t)$.

 One can write $K_1, K_2$ as follows
 \be K_1 +K_2 = \frac14  \sum_{i=1,2,}\int^{s_i}_0 d \tau_i \left(
 \frac{d\vez^{(i)}}{d\tau} \right)^2\label{28}\ee
 and introducing $\omega_i$ instead of $s_i, s_i =\frac{t}{2\omega_i}$ one
 obtains in the  exponent
\be K_1 +K_2  + V_{\rm conf} (\eta)  t \to\left(\frac{\vep^2_1}{2\omega_1} +
\frac{\vep^2_2}{2\omega_2} + \frac{\omega_1+\omega_2}{2} + V_{\rm conf}
(\eta)\right)t,\label{29}\ee where $\eta = |\vez^{(1)}-\vez^{(2)}|$.  On the
other hand one can introduce the unit operator
 \begin{eqnarray}
1  = 2\int ds_1 ds_2 \delta (s_1 +s_2 -s) \delta(s_1-s_2) =
 \nonumber\\ =
\int\frac{t d\omega_1}{\omega^2_1}\delta \left( \frac{t}{\omega_1} -s\right)
d\omega_2 \delta (\omega_2 -\omega_1)  = \frac{td\omega}{\omega^2} \delta\left(
\frac{t}{\omega} -s\right).\label{30}\end{eqnarray} Using Eq. (17) in
\cite{39*} one can rewrite (\ref{26}) with(\ref{30}) as

 \be G_3 (s)  = \int \frac{ td t  d \omega}{2\pi s \omega^2}\delta \left( \frac{t}{\omega } -s\right)
  d^2 u  \lan xx
 |e^{-H(\veP)t}|uu\ran,\label{31}\ee
where

  \be H (P)= \frac{\veP^2}{4\omega} + \frac{\vep^2}{\omega} +\omega+ V_{\rm conf},\label{32}\ee
  and finally, integrating out the free center-of-mass coordinate

\be \int d^2 u \lan xx|e^{-H(\veP)t}|uu\ran= \int d^2u \frac{d^2\veP}{(2\pi)^2}
e^{i\veP(\vex-\veu)} \lan 0| e^{-H(\veP)t}|0\ran=\lan 0
|e^{-H(0)t}|0\ran,\label{33}\ee where  in $\lan0|,|0\ran$,  enter only w.f. of
relative motion.

The eigenvalues of $H(0)$ can be found in the  same way, as it was done in
\cite{34}, using the local limit of $H(0)$ in $\omega$ at $\omega=\omega_0$,
\be M= 4 \omega^{(0)}_\nu;~~\omega^{(0)}_\nu =\left(
\frac{a_\nu}{3}\right)^{3/4} \sqrt{\sigma_{\rm adj}}, ~~\sigma_{\rm
adj}=\frac94\sigma_s, a_0 =1.74,\label{34}\ee which yields the lowest
eigenvalues $$\omega_0^{(0)} \approx \sqrt{\sigma_s}, ~~ M_0 = 4
\sqrt{\sigma_s}.$$ Finally one obtains \be G_3(s) =\frac{1}{\sqrt{\pi s}}
\sum_{\nu=0,1,...} \psi^2_\nu (0) e^{-M_\nu\omega_\nu^{(0)} s}\label{35}\ee and
$\psi^2_\nu (0) = c_\nu \sigma_s$, where the dimensionless constant $c_\nu$ has
to be defined, solving the wave equation with the Hamiltonian $H(0)$.

Hence the lowest mass squared in (\ref{35}) is \be \mu^2_0=M_0 \omega_0^{(0)}
\cong 4\sigma_s \approx m^2_D,\label{36}\ee where $m_D$ is the screening mass
found in \cite{22}. One can check the general expression (\ref{35}) in the free
case, $\sigma_s\equiv 0$. In this case $\sum_n \psi^2_n (0) =
\frac{d^2p}{(2\pi)^2}$ and $M_n, \omega_n^{(0)}$ from $H_0 =
\frac{\vep^2}{\omega} +\omega$, Eq. (\ref{32}), are $\omega_0 =|\vep|, M_0 =2p$
and one obtains the exact free result.

\be G_3^{(0)} (s)  = \frac{1}{\sqrt{\pi s}} \int \frac{d^2p}{(2\pi)^2}
e^{-2p^2s} = \frac{1}{\sqrt{\pi s}} \frac{1}{8\pi s} = \frac{1}{(4\pi
s)^{3/2}}, \label{37}\ee which using (\ref{15}) and (\ref{7}) yields the
Stefan-Boltzmann result $(L_{\rm adj}\equiv 1)$

\be P_{gl}^{(0)} = \frac{N^2_c-1}{(4\pi)^2} \int^\infty_0 \frac{ds}{s^3}
\sum_{n=\pm 1,\pm2} e^{-\frac{n^2}{4T^2s} } = \frac{2(N_c^2-1)T^4}{\pi^2}
\sum^\infty_{n=1} \frac{1}{n^4} = \frac{(N^2_c -1) T^4\pi^2}{45}.\label{38}\ee

Using (\ref{35}) one can write $P_{gl}^{(1)}$ as (keeping the only  term with
$\nu=0$, $\psi^2_0 (0) \equiv \bar c \sigma_s$)

\be P_{gl}^{(1)} = \frac{N^2_c -1}{(4\pi)^2} \int^\infty_0 \frac{ds}{s^2} \bar
c \sigma_s e^{- m^2_D (T) s} \sum_{n=\pm 1, \pm 2}
e^{-\frac{n^2}{4T^2s} }L_{\rm adj}^{(n)}.\label{39}\ee

From the integral representation of the modified  Bessel function \be K_\nu (z)
= \frac12 \left(\frac{z}{2} \right)^\nu \int^\infty_0
\frac{e^{-t-\frac{z^2}{4t}}}{t^{\nu+1}} dt,\label{40}\ee one arrives at the
following form (taking into account, that   $L_{\rm adj}^{(n)}\approx (L_{\rm
adj})^n $ for $T\la \lambda^{-1} =1$ GeV, as it is shown in \cite{10})\be
P_{gl}^{(1)} (T) =\frac{(N^2_c-1)\bar c \sigma_s m_DT}{2\pi^2} \sum_{n=1,2,...}
\frac{1}{n} K_1 \left( \frac{nm_D}{T}\right) (L_{\rm adj})^n.\label{41}\ee

On the other hand one can use the relation \be \sum_{n=1,2,..}
\frac{K_\nu(nz)}{n^\nu} = \frac{\sqrt{\pi}}{\Gamma \left( \nu+\frac12\right) (2
z )^\nu} \int^\infty_0 \frac{t^{2\nu} dt}{\sqrt{t^2+z^2}(\exp (\sqrt{t^2+z^2})
-1)},\label{42}\ee and one obtains \be P_{gl}^{(1)} (T) =\frac{(N^2_c-1)\bar c
\sigma_s  T^2 }{2\pi^2} \int^\infty_0 \frac{t^{2 } dt}{\sqrt{t^2 +
\left(\frac{m_D}{T}\right)^2}}\frac{1}{  \exp \left(\sqrt{t^2
+\left(\frac{m_D}{T}\right)^2}+a\right)-1},\label{43}\ee $ L_{\rm adj}=\exp
(-a)$.

Note, that we have kept the lowest eigenvalue $\nu=0$ in (\ref{35}), in a more
general case one should replace $\bar c \to \bar c_\nu$, $m_D \to m_D^{(\nu)}$
and sum over $\nu, \nu=0,1,2,...$ However, having in mind, that $m_D^{(\nu)}$
strongly rise in magnitude with growing $\nu$, and they enter in the exponent
in (\ref{43}), one can expect that the first term with $\nu=0$ yields a
reasonable approximation for not large $T$. In what follows we keep the form
(\ref{43}) with $\bar c$ being a free constant, to be fixed by comparison with
lattice data at some point of $T$.

One can simplify the answer in the case, when the  spatial confinement has the
form of  an oscillator potential. In this case one can write $G^{(n)} (s)$ in
(\ref{9}) as \be  G^{(n)} (s) = \int (Dz_4)^w_{0n} (Dz_3)_{00} (Dz_1)_{00}
(Dz_2)_{00} e^{-K} = \frac{1}{4\pi s} e^{-\frac{n^2}{4T^2s} } G_2
(0,0,s),\label{I}\ee \be G_2 (0,0, s) = \int (Dz_1)_{00} (Dz_2)_{00}
e^{-K_1-K_2} = \frac{M^2_0}{4\pi {\rm sh} M^2_0 s}.\label{II}\ee

Here $M_0=\omega$ is the lowest mass (excitation) in the oscillator potential,
which we might associate with the lowest screening mass $m_D$.

As a result one obtains the gluon pressure in the form \be P_{gl}^{(OCS)} =
\frac{2(N_c^2-1)}{(4\pi)^2}\sum^\infty_{n=1} L^{(n)}_{adj} \int^\infty_0
\frac{ds}{s^2} e^{-\frac{n^2}{4T^2s} }\frac{M^2_0}{  {\rm sh} M^2_0
s}.\label{III}\ee One can check, that for $M_0\ll T$ (\ref{III}) yields the
Stefan-Boltzmann result (\ref{38}), augmented by the term $L^n$.

To make a connection with the realistic case of linear confinement, $V(r) =
\sigma_s r$, one can make a substitution $\sigma_s r \to \frac{\sigma_s}{2}
\left( \frac{r^2}{\gamma} + \gamma\right)$, which after variation in the
parameter  $\gamma$ yields back the linear potential. The use of this trick was
checked to give approximately 5\% accuracy in the  spectrum calculations. As a
result one obtains  a crude  approximation for $G_3 (s)$, Eq. (\ref{16})  of
the linear potential

\be G_3^{\rm lin} (s) \to \frac12 (\gamma G_3^{(0)} (s) + \frac{1}{\gamma}
G_3^{(OSC)} (s) \to \frac{1}{(4\pi s)^{3/2}} \sqrt{\frac{M^2_0 s}{{\rm sh} M^2_0
s}}\label{47*}\ee and  as  a result one obtains \be P_{gl} =
\frac{2(N_c^2-1)}{(4\pi)^2} \sum^\infty_{n=1} L^{(n)}_{adj}\int^\infty_0
\frac{ds}{s^3} e^{-\frac{n^2}{4T^2s} }\sqrt{\frac{M^2_0 s}{{\rm sh} M^2_0
s}}.\label{47**}\ee

In what follows we shall use (\ref{47**}) with $M_0\approx m_D$,  and  we shall
find  that  the  results of (\ref{III}) and (\ref{47**}) are rather close
numerically.

\section{Polyakov lines in the Field correlator approach}

As was discussed in the Introduction,  the CE gluon correlators produce the
potential $  V^{\rm sat}_{1}(r) = V_1(\infty) + v(r)$, Eq. (\ref{4b}), so that
in  the $gg$ Green's function acquires the factor  $\Lambda \equiv\exp \left(
-c_a \frac{V_1(\infty)}{2 }t_4\right)$ for each gluon, when one considers
$v(r)$ as a perturbation.

However, in the confined region $ V^{\rm sat}_{1}$ is screened by the  $V_D
(r,T)$, and therefore this factor $\Lambda$ appears only in the  deconfined
phase, where it appears in the form of the Polyakov line.

 In the Matsubara
representation of the temperature Green's function $G^{(n)}(s)$, Eq. (\ref{8}),
one has   the phase $J^E_n, $ Eq. (\ref{14}) which tends to  $
\frac{nV_1(\infty)}{2T }$ for $T\to 0$, in agreement with $\Lambda$, when
$t_4=1/T$.

Thus Eq. (\ref{14}) defines the Polyakov loop at $T>0$ and also at $T>T_c$ via
$V_1(r, T),$ namely

\be L_{\rm adj}^{(n)} =\exp \left( - \frac{9n}{8T} V_1^{(n)}(\infty,
T)\right).\label{47b}\ee \be V_1^{(n)} (\infty,T) = \int^{n/T}_0 d\nu
\left(1-\frac{\nu T}{n}\right) \int^\infty_0 \xi d\xi D_1^E
\left(\sqrt{\xi^2+\nu^2}\right).\label{47c}\ee

The important property to be used in what follows, is the short distance
behavior of $D_1^E(x)$, which is  concentrated at distances $|x| \la \lambda
=0.2$ fm and is assumingly not affected by $T$ for $T< 1/\lambda  \cong 1$ GeV
\cite{17,18,24*}. In this case one can make a replacement, $n \to 1$ in
(\ref{47c}), and omit the superscript $n$ in $V_1^{(n)} (r,T)$, as we shall do
in what follows writing \be L_{\rm adj}^{(n)} = (L_{\rm adj}(T))^n, ~~ L_{\rm
adj} (T) = \exp \left( - \frac{9V_1(\infty,T)}{8T} \right), \label{47d}\ee
where $V_1(\infty,T)$ is given in (\ref{47c}) via the correlator $D_1^E (x)$.

 Note two important consequences of our theory for $L(T)$: first of all the $Z(3)$
 symmetry of the SU(3) theory is spontaneously broken by the vacuum field
 correlator, which fixes one of 3 branches with $N=0$.

 Secondly, the Casimir scaling for $L_J(T)$  observed on the lattice \cite{19},
  appears naturally, since $V_1^{(a)}
 (T)$ is proportional to $c_a$.

 To compute $P(T), I(T)$ etc. numerically we need the explicit form of $V_1
 (\infty, T)$ or $D_1(x-y)$. In the phase II for  $T<T_c$ this can be derived, using the
 gluelump Green's  functions and  eigenvalues \cite{17}. Using the same form also for
 $T>T_c$ it was found in \cite{18}, that the function $V_1(\infty, T)$,  agrees
 approximately with the lattice free energy $F_1 (\infty, T)$ \cite{39a}.  In what
 follows we shall use this form, however we shall take into account that on
 general grounds $F_1 (\infty, T)< V_1(\infty, T)$ and  the  negative values of
 $F_1 (\infty, T)$ for large $T\gg T_c$ do not provide negative $V_1(\infty,
 T)$ and hence $L(T) \leq 1$ \cite{18}. One can also argue, that our $L(T) <
 L_{\rm lat }(T)$.

  The Polyakov line  can  also be  obtained from the  gluelump form of the correlator $D_1$, which can be written
  according to \cite{18} as

  \be
 D_1^{(np)}
(x) = \frac{A_1}{|x|} e^{-M_1|x|}+O(\alpha_s^2),~A_1=2C_2\alpha_s\sigma_{\rm
adj} M_1,~ x\geq 1/M_1.\label{47}\ee and the nonperturbative part of $V_1$
(note that $D_1$ contains also the perturbative gluon exchange correlator),
which for $T=0$ has  the form (\ref{1a}), for $T>0$ can be written as \be
V_1^{(np)} (r,T) =A_1\int_0^{1/T}(1-\nu T) d\nu \int^r_0 \frac{\xi d\xi
e^{-M_1\sqrt{\xi^2+\nu^2}}}{\sqrt{\xi^2+\nu^2}},\label{48}\ee which yields at $
r\to \infty $

\be L_f=\exp \left(-\frac{V_1^{(np)}(\infty)}{2T}\right),~~V_1^{(np)}(\infty)
=\frac{A_1}{M_1^2} \left[1-\frac{T}{M_1} (1-e^{-M_1/T})\right].\label{49}\ee

One can also    use directly its lattice renormalized
 values, and we shall prefer the
 $L_{ }^{\rm ren}(T)$ from \cite{19}, where $L^{\rm ren}_a (T)$ were found for
 different SU(3) representations  $a$, and the Casimir scaling was established
 with good accuracy.

The comparison of $L_{\rm adj}(T)$ in the region $T>T_c$ with the lattice data
\cite{19}  in Fig. \ref{fig:fig03} shows a  reasonable  agreement.

 One can compare  (\ref{48}), (\ref{49})  with  the lattice data for the
pair free energy $F_1$ \cite{39a}, which yields for  $V_1 (\infty, T)$   the
value $V_1(\infty, T_c)\approx 0.5$ GeV (with 10\% accuracy) and decreasing
with growing $T$, and we approximate $V_1^F(\infty, T)$ as obtained from $F_1$
\be V_1^F(\infty, T) = \frac{0.175~{\rm GeV}}{1.35 \frac{T}{T_c} -1} , ~~ T\geq
T_c\label{50}\ee

At this point one should stress, as it was also done in \cite{18} the
difference between $V_1(r,T)$ and $F_1(r,T)$, measured on the lattice, which
can be written as \be e^{-F_1(r,T)/T} = \sum_n e^{-E_n (r,t)/T}, \label{51}\ee
and $E_0 (r,T)$ can be associated with $V_1(r,T)$, while higher in $n$ states
make $F_1$ smaller than $V_1$, and finally can  make it negative at larger $T$,
as it was found on the lattice.

To  account for the difference $V_1$ and $F_1$ we can use another form
$V_1^{(\rm mod)}(\infty, T)$, where $V_1^{(\rm mod)}> F_1$, namely

\be V_1^{(\rm mod)} (\infty, T)=\frac{0.13{\rm~
GeV}}{T/T_c-0.84}.\label{52*}\ee

{
\begin{figure}[htb] 
\setlength{\unitlength}{1.0cm}
\centering
\begin{picture}(9.0,6.0)
\put(0.5,0.5){\includegraphics[height=5.5cm]{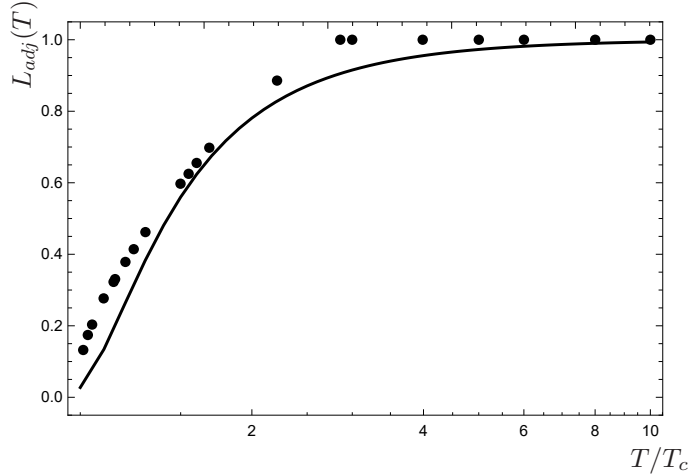}}
\put(8.25,0.1){\footnotesize  $T/T_c$}
\put(0.1,5.10){\rotatebox{90}{\footnotesize $L_{adj}(T)$}}
\end{picture}
\caption{Polyakov line $L_{\rm adj} (T)$: the solid line is our modified $L_{\rm adj}^{(\rm mod)}(T)$ from Eqs. (\ref{52*}),(\ref{52**}) and filled dots are for the lattice data \cite{19}.}
\label{fig:fig03}
\end{figure}
}\medskip

In Fig. \ref{fig:fig03} we show  both the lattice data for $L_{\rm adj} (T)$
taken from \cite{19} and our modified $L_{\rm adj}^{(\rm mod)}(T)$, calculated
as \be L_{\rm adj}^{(\rm mod)} (T) =\exp \left( -\frac{9V_1^{(\rm
mod)}(\infty,T)}{8T}\right).\label{52**}\ee One can see a reasonable agreement
between two lines, satisfying the required relation $L_{\rm adj}^{(\rm
mod)}(T)\la L_{\rm adj}^{\rm lat }(T)$.  The form (\ref{52**}) is used below in
our calculations of all thermodynamic functions.

{
\begin{figure}[htb] 
\setlength{\unitlength}{1.0cm}
\centering
\begin{picture}(8.0,4.1)
\put(0.5,0.5){\includegraphics[height=3.65cm]{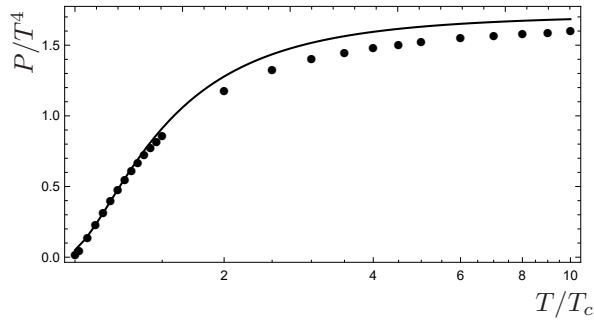}}
\put(7.0,0.1){\footnotesize $T/T_c$}
\put(0.1,3.25){\footnotesize \rotatebox{90}{$P/T^4$}}
\end{picture}
\caption{The pressure ${P(T)}/{T^4}$ in the $SU(3)$ theory in the deconfined phase. The solid line is for the modified oscillator confinement Eq. (\ref{47**}), and filled dots are for the lattice data  \cite{43}.}
\label{fig:fig04}
\end{figure}
}\medskip

The resulting pressure $P_{gl}(T)$ for $T\geq T_c$  is shown in Fig.
\ref{fig:fig04}. One can see, that the use of $L_{\rm adj}^{(\rm mod)}(T)$ from
(\ref{52**}), (\ref{52*}) and   of the magnetic confinement, Eq. (\ref{47**})
gives    a reasonable agreement with lattice $SU(3)$ data from \cite{43}.

 \section{ The confinement  sector}


We now turn to the confined  gluonic phase, which  consists of the two-gluon,
three-gluon, etc. glueballs, which can be calculated analytically via
$\sigma^{(E)} $ \cite{31}. The corresponding pressure of the noninteracting gas
of glueballs of the i-th kind with mass $m_i$ is \cite{32} \be P_{gb}^{(i)} =
\frac{g_iT^2}{2\pi^2} \sum^\infty_{n=1}  \frac{ m_i^2}{n^2} K_2 (\frac{n m_i}{T}),
\label{20}\ee where $g_i$ is the multiplicity of the  i-th glueballs.

We have disregarded in (\ref{20}) the contribution of the possible real or
virtual glueball decay products, as well as the interaction between glueballs,
which disappears in the large $N_c$ limit.


 The total  pressure, $P_{\rm conf}$,   in the SU(3) case is given
 by  the sum of  the glueball terms (\ref{20}), namely
 \be P_{\rm conf} = \sum_i P^{(i)}_{gb}.\label{54}\ee

 The situation here depends on the spectrum of lowest glueballs, which was
 found  repeatedly on the lattice \cite{38,39,40} and also  analytically in the Field
 Correlator Method \cite{31}, see comparison in the Table 1, which shows a remarkable agreement of almost all states.
 One expects that, the
 total contribution of the excited glueballs might   be important in the
 region near $T_c$,   and  the question arises, how one approximates the asymptotic
 behavior of the spectrum.

 A most detailed  lattice analysis of the SU(3) thermodynamics done recently in
 \cite{43}, reveals that e.g. the trace anomaly below and near $T_c$ can be
 described by a combination of glueball and Hagedorn contributions \cite{43a} (see Figs. 3
 and 4 in \cite{43}).

 In an accurate analysis of the entropy density $s$  in \cite{44} it was found,
 that $0^{++}$ and $ 2^{++}$ glueballs contribute to $s/T^3$ less than 25\%
 at $T=T_c$, and only the  combination of glueballs with mass less than $2M_0$
 (two-particle threshold) and the  Hagedorn spectrum
 corresponds to the lattice data.

However, from our point of view the use of the Hagedorn density of states
\cite{43a}, derived from  the closed string spectrum, in addition to the 10-12
lowest glueballs, seems to be  superfluous. Indeed, there is no evidence that
the Hagedorn spectrum has  quantitative correspondence with the realistic
glueball spectrum, and that  the closed strings are  resemblant to multigluon
glueball states. One can also add, that strickly speaking  the 4d string theory
does not exist.

Moreover, it is hard to imagine, that high excited closed string states are
realized on the finite size lattice.

Therefore we turn to another explanation of the high growing glueball
contribution to $P_{\rm  conf}$ near $T_c$. Namely, it was repeatedly found on
the lattice (see e.g. \cite{29*}, \cite{29**} and \cite{29***}), that the
string tension $\sigma_E$ starts to depend on $T$ in the region $0.7 T_c\leq T
\leq T_c$, and tends to a value $\sigma_E(T_c)$, which is in the region $0.2
\sigma_0 \leq \sigma_E (T_c) \leq 0.5\sigma_0$. here $\sigma_0 = \sigma_E
(T=0)$. It is clear physically, that glueball masses decrease   as $m_i (T) = a
(T) m_i (0)$, where $a(T) = \sqrt{\frac{\sigma_E (T)}{\sigma_0}}$.

As a result in (\ref{20}) one obtains a strong amplification of the glueball
pressure. Indeed, writing $a(T)$ as \be a(T) = \sqrt{1-\left(
\frac{T}{T_c+b}\right)^2}\label{**}\ee one obtains the pressure $P_{\rm conf} $
in (\ref{54}) for 12  and 2 lowest glueballs, shown in Fig. \ref{fig:fig05}.

\begin{table}[!htb]
\caption{{Glueball  masses  from FCM as compared to lattice data}}
 \begin{center}
\label{tab.01}\begin{tabular}{|c|c|c|c|c|  }\hline

$J^{PC}$&$M$(GeV)&\multicolumn{3}{c|}{Lattice data}\\ \hline

&Ref. [41]& Ref. [51]& Ref. [52] & Ref. [53]\\\hline

$0^{++}$&1.58&1.710(50)(80)& $1.73\pm 0.13$ &$1.74\pm 0.05$\\\hline

$0^{++*}$ & 2.71& &2.67$\pm 0.31$& $3.14\pm 0.10$\\ \hline

$2^{++}$ & 2.59&2.39 &2.40$\pm 0.13$& $2.47\pm 0.08$\\ \hline

$2^{++*}$ & 3.73& &3.29$\pm 0.16$& $3.21\pm 0.35$\\ \hline

$0^{-+}$ & 2.56&2.56 &2.59$\pm 0.17$& $2.37\pm 0.27$\\ \hline

$0^{-+*}$ & 3.77& &3.64$\pm 0.24$& \\ \hline

$2^{-+}$ & 3.03&3.04 &3.1$\pm 0.18$& $3.37\pm 0.31$\\ \hline

$2^{-+*}$ & 4.15& &3.89$\pm 0.23$&\\ \hline

$3^{++}$ & 3.58& 3.67&3.69$\pm 0.22$& $4.3\pm 0.34$\\ \hline

$1^{--}$ & 3.49& 3.83&3.85$\pm 0.24$& \\ \hline

$2^{--}$ & 3.71&4.01 &3.93$\pm 0.23$& \\ \hline

$3^{--}$ & 4.03&4.20 &4.13$\pm 0.29$& \\ \hline
\end{tabular}

\end{center}
\end{table}

 One can see in Fig. \ref{fig:fig05} the resulting $P_{\rm conf}(T)$  as a function of $T$ in
 comparison with the  lattice data \cite{43} for two cases: 1) when only
 $0^{++}$ and $2^{++}$ glueballs are retained, and   2), when 12 lowest
 glueball states are included with $a(T)$ (\ref{**}) and $b=0.15~ T_c$. One  can see  a good agreement  of
  $P_{conf}(T)$ in the case 2) with the lattice data from  \cite{43} for  the    chosen  value of  $b$.
   At the same time we present in Fig. \ref{fig:fig06} the  comparison  of our
   resulting behavior   of $\frac{\sigma (T)}{\sigma_0}$ with the  lattice
   measurements  of confinement    attenuation in \cite{29*,29**,29***}, which shows  a
    reasonable  qualitative agreement.

Thus we conclude, that there is no need to exploit the Hagedorn mechanism for
the explanation of the pressure $P_{conf}$ near $T_c$.

{
\begin{figure}[htb] 
\setlength{\unitlength}{1.0cm}
\centering
\begin{picture}(8.0,4.0)
\put(0.5,0.5){\includegraphics[height=3.55cm]{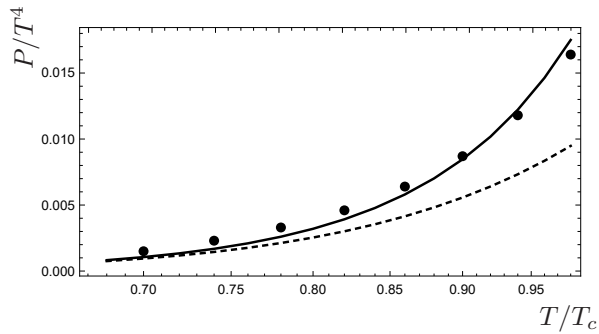}}
\put(7.05,0.1){\footnotesize $T/T_c$}
\put(0.1,3.5){\footnotesize \rotatebox{90}{$P/T^4$}}
\end{picture}
\caption{Pressure in the confining phase. The dashed line is for 2 lowest glueballs ($0^{++}$ and $2^{++}$) and the solid line is for 12 glueballs. The filled dots are for the lattice data \cite{43}.}
\label{fig:fig05}
\end{figure}
}\medskip

{
\begin{figure}[htb] 
\setlength{\unitlength}{1.0cm}
\centering
\begin{picture}(7,5.0)
\put(0.35,0.15){\includegraphics[height=4.25cm]{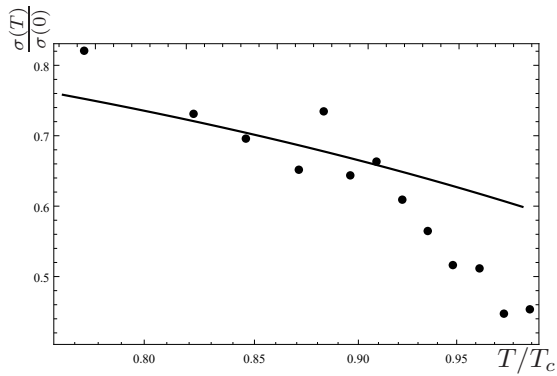}}
\put(6.45,0.1){\footnotesize $T/T_c$}
\put(0.025,4.25){\footnotesize \rotatebox{90}{$\frac{\sigma(T)}{\sigma(0)}$}}
\end{picture}
\caption{The solid line is for the string tension $\sigma(T)/\sigma(0)$ calculated from Eq. (\ref{**}), and dots are for the lattice data \cite{29**}.}
\label{fig:fig06}
\end{figure}
}\medskip

\section{ Results for the SU(3) phase transition and trace anomaly}

In this section we combine together our results for the confined and deconfined
phases. In doing so we calculate also the trace anomaly $\frac{I(T)}{T^4}
=\frac{\varepsilon-3P}{T^4}$, and  the entropy density $s(T) = \left(
\frac{dP(T)}{dT}\right) \frac{1}{T^3}$.

We calculate $P_{gl} (T)$, as in (\ref{47**}) with the account of the Polyakov
loops $L_{\rm adj} (T)$, given in (\ref{52*}),(\ref{52**}) and the
colormagnetic confinement as in (\ref{47**}).  The comparison of our $P_{gl}
(T)$ and the corresponding lattice  values from \cite{43} in Fig.
\ref{fig:fig04} shows a good agreement in the interval $T_c\leq T \leq 10
~T_c$. For  $P_{\rm conf}$ eqs. (\ref{20}) and (\ref{54}) are used with masses
$m_i (T) = a(T) m_i(0)$, where $a(T)$ is given in  (\ref{**}) and  masses
$m_i(0)$ in Table 1, the first column.

In $P_{\rm conf}$ we distinguish two cases with number of glueballs equal to a)
2 and b) 12, and $a(T)$ given in (\ref{**}). These analytic results are shown
in Fig. \ref{fig:fig05} in comparison with lattice data  from \cite{43}.

 We are using the phase transition  condition,
which can be written as \be P_{gl} (T_c) =   P_{\rm conf} (T_c),\label{64}\ee
which yields $T_c \simeq 260$ MeV, as shown in Fig. \ref{fig:fig07}. This
agrees with lattice data from \cite{19*,16*,1,2} and \cite{43}.

 An important measure of the interaction is the trace anomaly, which we compute
 analytically as $I(T)  = \varepsilon - 3p$ both below $T_c$ in Fig. \ref{fig:fig08}  and
 above $T_c$ in Fig. \ref{fig:fig09}.  The results for $\frac{I(T)}{T^4}$ are compared with
 the lattice data from \cite{43} and demonstrate a good agreement.

As a next  step    we find $I_{<}(T_c)$  from the confinement phase and
$I_{>}(T_c)$ for the deconfined phase and calculate the difference $\frac{\Delta
I(T_c)}{T^4} = \frac{I_{>}(T_c)}{T^4}-\frac{I_{<}(T_c)}{T^4}$, which for $T_c=0.260$ GeV is equal to $\frac{\Delta
I(T_c)}{T^4}=0.61$, while $\frac{\Delta
\varepsilon(T_c)}{T^4}=0.66$.

One can compare this value with the lattice data from \cite{56}, $\frac{\Delta
(\varepsilon - 3P)}{T_c^4} = 0.6223\pm 0.056$, while in \cite{44} it was
obtained $\frac{\Delta (\varepsilon - 3P)}{T_c^4} =1.39 (4)(5)$. This latter
value  is close to the  measured in \cite{19*}  and \cite{57}.

We now turn to the behavior of $I(T)$ for $T>T_c$, where the lattice data
\cite{43} discovered an interesting shoulder in the dependence of
$\frac{I(T)}{T^2T^2_c}$ in the range $T_c\leq T\leq 4 T_c$.

It was shown in our previous work \cite{39***}, that this is of purely np
origin and is provided by $1/T^2$ behavior of $L(T)$. One can see our analytic
results in reasonable agreement with the lattice data for $\frac{I(T)}{T^4}$
and $\frac{I(T)}{T^2 T^2_c}$ in Figs.  \ref{fig:fig09}, \ref{fig:fig10}.

Finally in Fig. \ref{fig:fig11} we show the entropy density
$\frac{s(T)}{T^3}$, which
  agrees with  lattice data from \cite{43}.

{
\begin{figure}[htb] 
\setlength{\unitlength}{1.0cm}
\centering
\begin{picture}(9.0,6.5)
\put(0.3,0.5){\includegraphics[height=5.0cm]{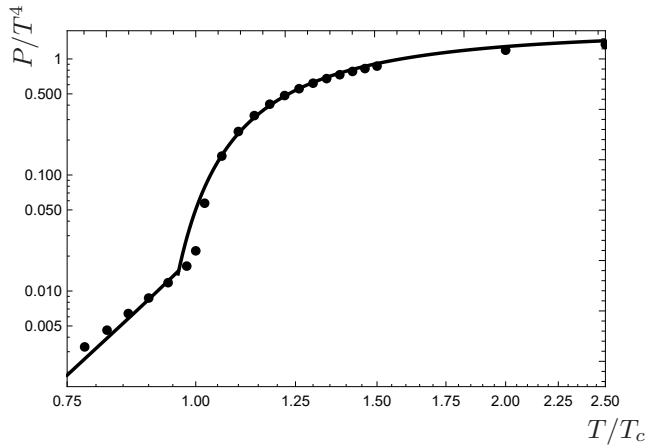}}
\put(7.70,0.1){\footnotesize $T/T_c$}
\put(0.1,5.0){\footnotesize \rotatebox{90}{$P/T^4$}}
\end{picture}
\caption{The pressure of the $SU(3)$ theory in the confining and deconfining phases. Filled dots are for the lattice data \cite{43}.}
\label{fig:fig07}
\end{figure}
}\medskip

{
\begin{figure}[htb] 
\setlength{\unitlength}{1.0cm}
\centering
\begin{picture}(9.0,4.85)
\put(0.5,0.5){\includegraphics[height=4.0cm]{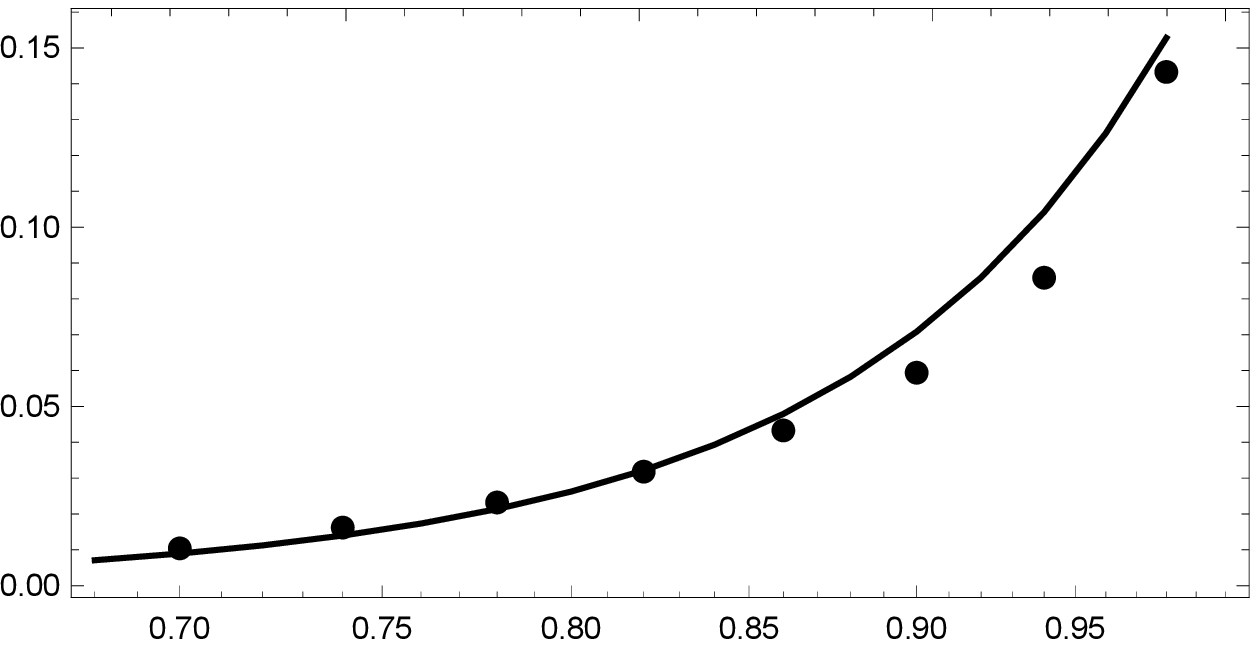}}
\put(7.85,0.1){\footnotesize $T/T_c$}
\put(0.1,3.9){\footnotesize \rotatebox{90}{$I/T^4$}}
\end{picture}
\caption{The trace anomaly in the confined phase, filled dots are for the lattice data \cite{43}.}
\label{fig:fig08}
\end{figure}
}\medskip

{
\begin{figure}[htb] 
\setlength{\unitlength}{1.0cm}
\centering
\begin{picture}(9.0,4.85)
\put(0.5,0.5){\includegraphics[height=4.0cm]{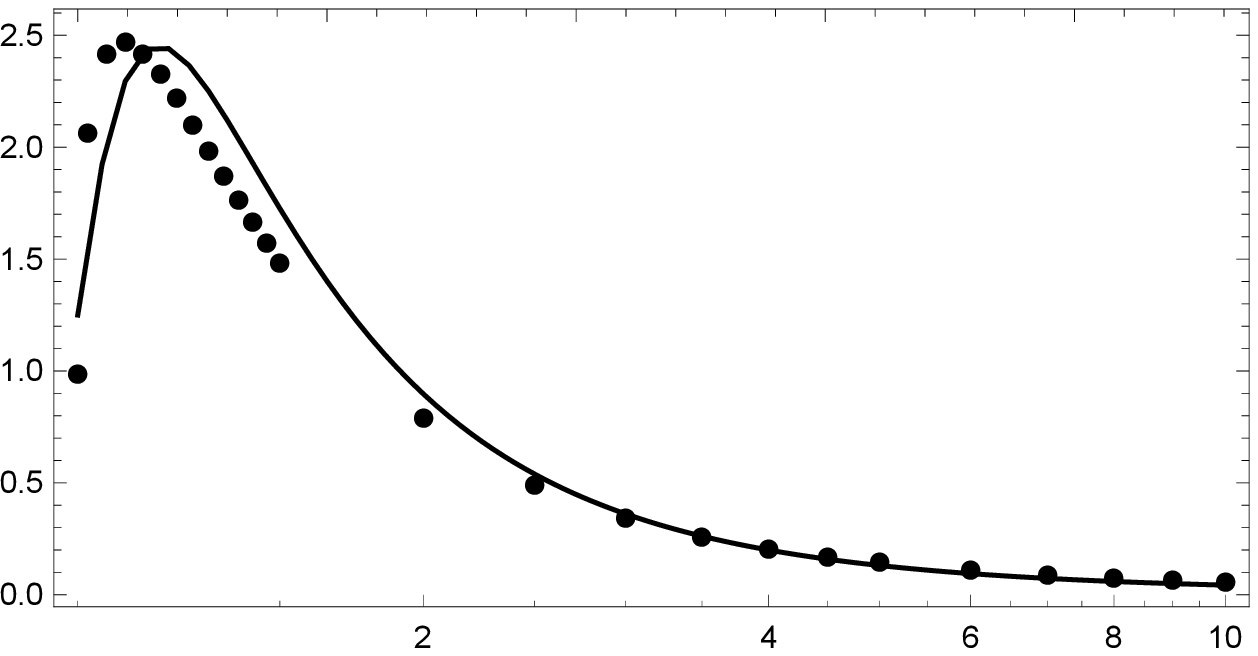}}
\put(7.85,0.1){\footnotesize $T/T_c$}
\put(0.1,3.9){\footnotesize \rotatebox{90}{$I/T^4$}}
\end{picture}
\caption{The trace anomaly in the deconfined phase, filled dots are for the lattice data \cite{43}.}
\label{fig:fig09}
\end{figure}
}\medskip
\vspace{0.5cm}

{
\begin{figure}[h] 
\setlength{\unitlength}{1.0cm}
\centering
\begin{minipage}[h]{0.485\linewidth}
\begin{picture}(6.0,4.5)
\put(0.75,0.3){\includegraphics[height=3.45cm]{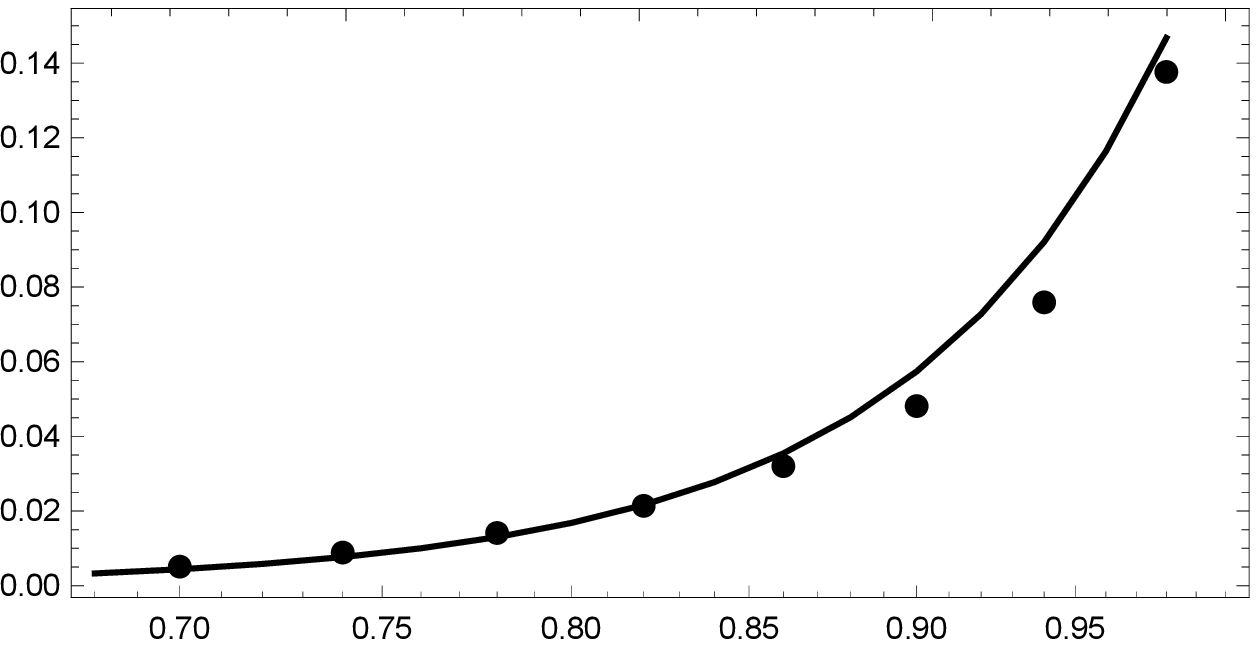}} 
\put(6.95,0.1){\scriptsize $T/T_c$}
\put(0.1,3.2){\scriptsize  \rotatebox{90}{$\frac{I}{T^4}\left( \frac{T}{T_c} \right)^2$}}
\end{picture}
\end{minipage}
\hfill
\begin{minipage}[h]{0.485\linewidth}
\begin{picture}(6.0,4.5)
\put(0.75,0.3){\includegraphics[height=3.45cm]{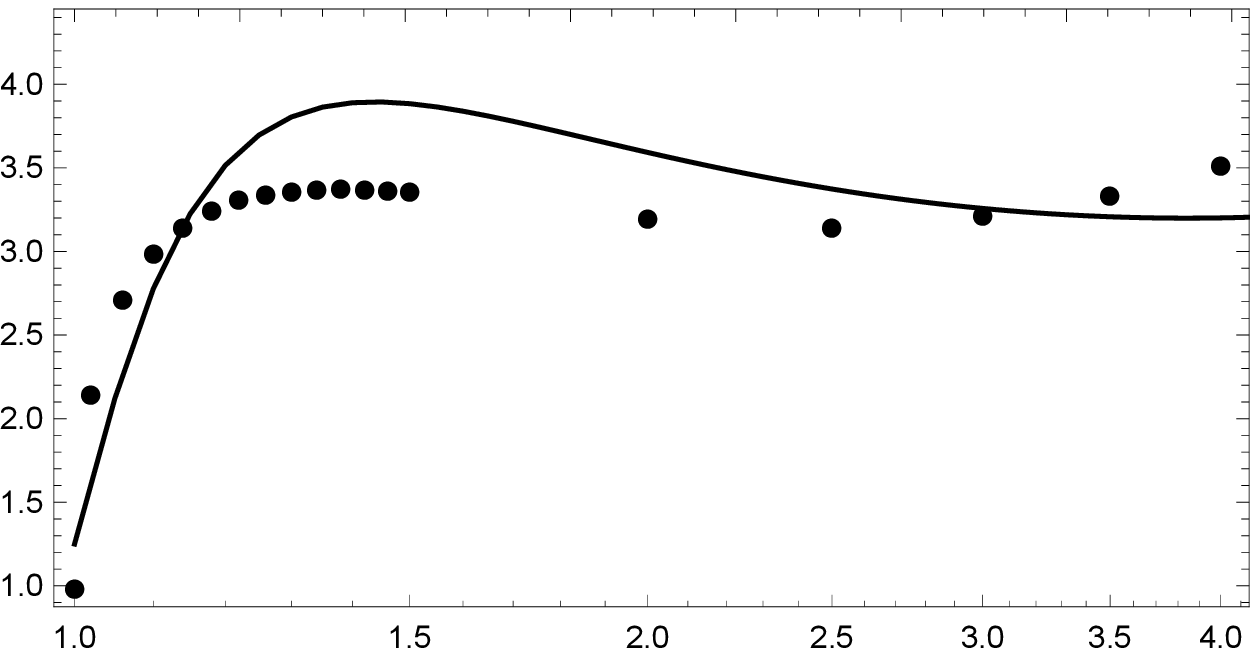}} 
\put(6.95,0.1){\scriptsize $T/T_c$}
\put(0.1,3.2){\scriptsize \rotatebox{90}{$\frac{I}{T^4}\left( \frac{T}{T_c} \right)^2$}}
\end{picture}
\end{minipage}
\caption{The trace anomaly multiplied by $(T/T_{c})^2$ in the confined phase --- the left panel, and in the deconfined phase --- the right panel. The filled dots are for the lattice data \cite{43}.}
\label{fig:fig10}
\end{figure}
}\medskip

{
\begin{figure}[h] 
\setlength{\unitlength}{1.0cm}
\centering
\begin{minipage}[h]{0.485\linewidth}
\begin{picture}(6.0,4.0)
\put(0.2,0.3){\includegraphics[height=3.45cm]{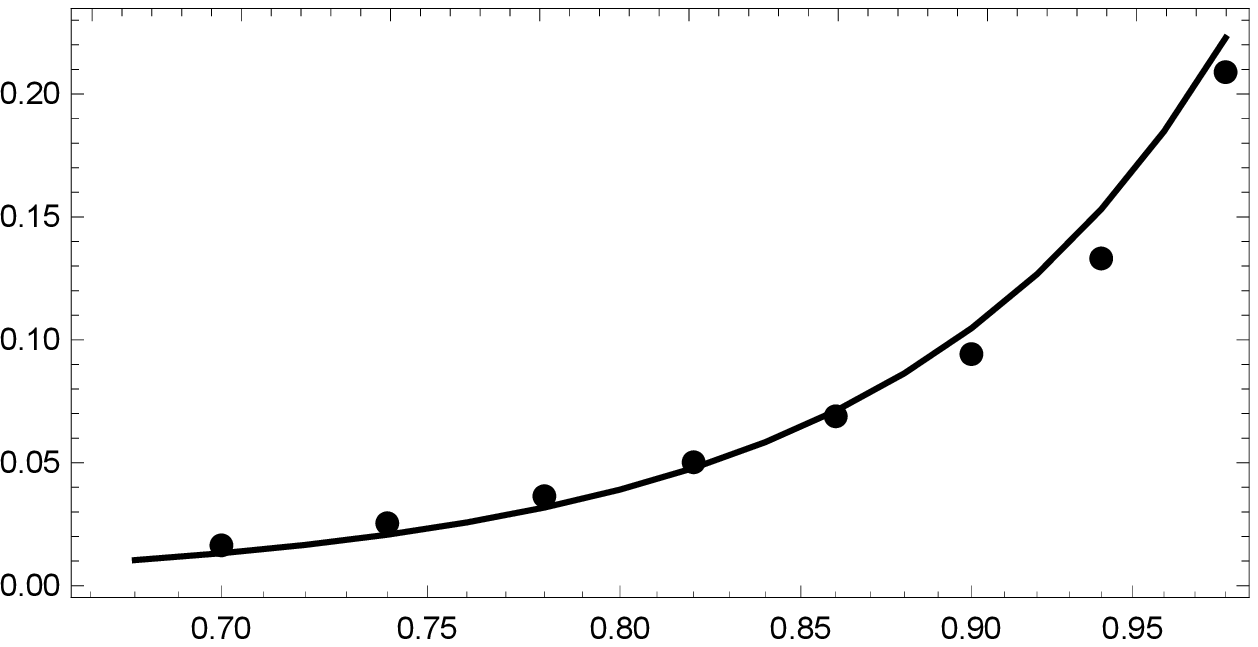}} 
\put(6.75,0.1){\scriptsize $T/T_c$}
\put(0.1,3.5){\scriptsize \rotatebox{90}{$s/T^3$}}
\end{picture}
\end{minipage}
\hfill
\begin{minipage}[h]{0.485\linewidth}
\begin{picture}(6.0,4.0)
\put(0.4,0.3){\includegraphics[height=3.45cm]{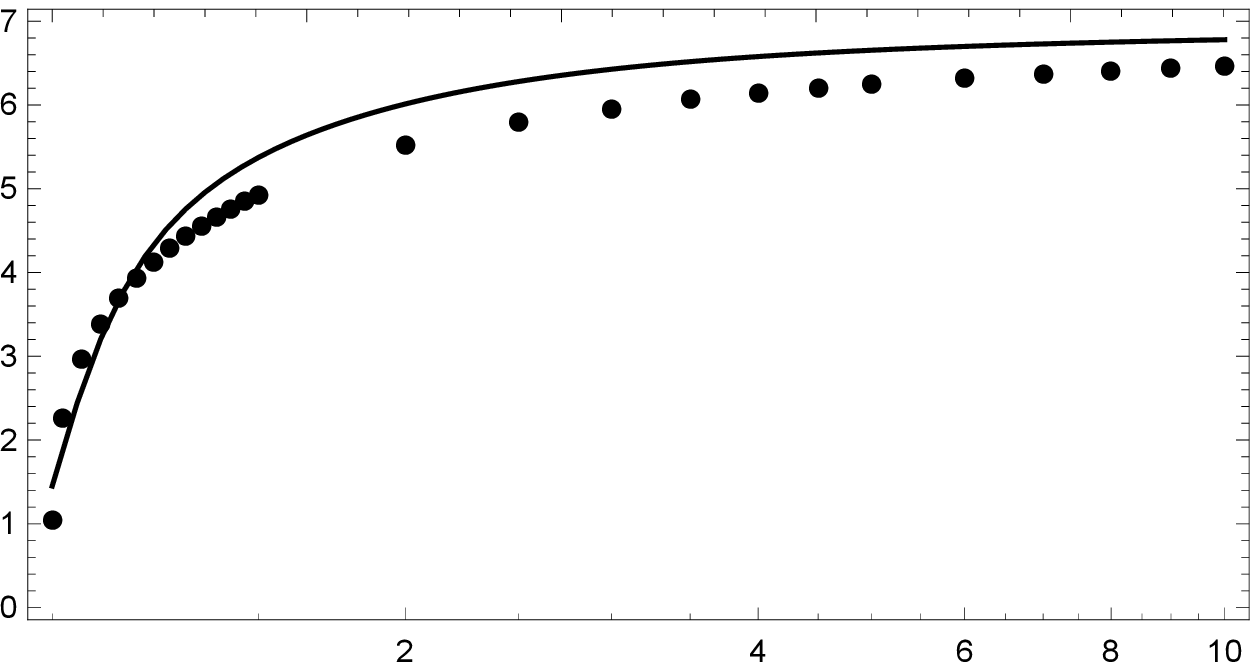}} 
\put(6.75,0.1){\scriptsize $T/T_c$}
\put(0.1,3.5){\scriptsize \rotatebox{90}{$s/T^3$}}
\end{picture}
\end{minipage}
\caption{The same as in Fig. \ref{fig:fig10} but for the entropy density.}
\label{fig:fig11}
\end{figure}
}\medskip


\section{Discussion of results and conclusions}

In this paper, as well as in our previous paper \cite{39***}, we have used the
standard definition of the pressure $P(T)$ and other thermodynamic
characteristics, both below and above $T_c$, without including in $P(T)$ vacuum
contributions $\Delta \varepsilon_{vac} V_3$, as it was done  in the previous
papers \cite{8,9,10,11*,11}. This has allowed us to make a direct  comparison
of our analytic and numerical results with other approaches and first of all,
with the numerical results of lattice calculations. The   accurate lattice data
of \cite{43}  for $P(T), I(T)$ and $s(T)$ have been used to  compare with our
results, which  demonstrates a satisfactory to  a  good  agreement between the
corresponding data.

We have kept in the present paper the same approach, as in previous ones, of
the explicit definition of two phases with two different dynamics:  the
confined phase with CE and CM confinement and correlators, and the  suppressed
Polyakov lines, and the deconfined phase with CM confinement and correlators
and resurrected  Polyakov lines.

We have used confining interaction, derived and checked numerously to calculate
lowest glueball masses in good  agreement with lattice data, to calculate
$P_{\rm conf} (T)$. In doing so, we have applied the variable vacuum principle,
allowing to suppress vacuum contribution to  the dynamics (e.g. the string
tension $\sigma (T)$), if it results in the increasing of $P(T)$.

In this way $\sigma (T)$ decreases for $T\ga 0.7 T_c$,, making the glueball
masses lighter and enhancing $P(T)$ in good agreement with numerical lattice
data from \cite{43}.

The effect of the temperature dependence of the string tension $\sigma (T)$ is
well known from numerous lattice measurements, see e.g. \cite{29*,29**,29***},
which support the principle mentioned above.

The comparison of our curves for $\sigma (T)$ with the lattice  data from
\cite{29*,29**,29***} in Fig. \ref{fig:fig06} shows a qualitative agreement.

This point has allowed to avoid the use of the popular Hagedorn fitting, which
is not well founded in our case, as we stressed above. Moreover, the latter is
not exploited in the case of $n_f>0$.

However our form of the string tension quenching, Eq.(\ref{20}) is still the
fitting procedure. It agrees qualitatively with the lattice data, as shown in
Fig. \ref{fig:fig06}, but should  be  derived analytically, and  this  work is
planned for the  future.

For $T>T_c$ we are using two main dynamical effects, the Polyakov loops $L_{\rm
adj} (T)$, which are shown to enter linearly in $P(T)$, and CM  confinement
yielding CM screening mass, and reducing the pressure from the upper limit of
the Stefan-Boltzmann law. In doing so we are using the slightly higher Debye
mass, $ M_0 \simeq 2 m_D \simeq 4\sqrt{\sigma_s}$,  however the results for
the proper value of $m_D$ do not differ much. For Polyakov lines $L_{\rm adj}
(T)$ we are using equations (\ref{52*}), (\ref{52**}), which are close both to the analytic forms obtained
earlier in Eqs. (\ref{49}), (\ref{50}), and  to the lattice data from \cite{19}.

With these modest input data we have obtained   results for $P(T), I(T)$ and
$s(T)$, which are shown in Figs. \ref{fig:fig05}-\ref{fig:fig11}, demonstrating
a good agreement with the lattice data \cite{43}.

The same is true for the value of $T_c \simeq 260$ MeV, found from Fig.
\ref{fig:fig07}. Summarizing, one can say, that the confining and nonconfining
dynamics considered here, is  supported by independent numerical data, and can
be used to develop further our approach in application to the real QCD $(n_f =
2+1)$, as well as to the interesting cases of $n_f =2$ and arbitrary $N_c$. The
work  of M.S.L. and Yu.A.S. was done in the framework of the scientific
program of the Russian Science Foundation, RSF, project 16-12-10414.

\vspace{2cm}

{\bf Appendix  }\\

{\bf  The $V_1$ cancellation  in the confinement region }\\

 \setcounter{equation}{0} \def\theequation{A.\arabic{equation}}

 As it was shown in (\ref{5b}), the instantaneous $q\bar q$ interaction can be written
 as
 \be V_{q\bar q}(r) = V_{\rm lin} (r) + \bar V_{\rm sat} (r), \label{A1}\ee
 where
 \be V_{\rm lin} (r) = 2 r \int^r_0 d\lambda \int^\infty_0 d\nu D^E (\lambda,
 \nu),\label{A2}\ee
 and the saturated at large $r$ potential $\bar V_{\rm sat} (r)$  is
 \be \bar V_{\rm sat} (r)= \int^r_0\lambda d\lambda \int^\infty_0 d\nu  [D^E_1 (\lambda,
 \nu)-2D^E  (\lambda,
 \nu)].\label{A3}\ee

 In what follows we show, that    $\bar V_{\rm sat} (r)$ is strongly suppressed in the
 confining region due to cancellation of $D_1^E$ and $D^E$, while  it is equal to $V_{1} (r) \equiv   \int^r_0\lambda d\lambda \int^\infty_0 d\nu D^E_1 (\lambda,
 \nu)$ in the deconfined region, and $V_1(r) = V_1^{(np)} + V_1^{\rm pert}$.

 To this end one can use the gluelump representation of the correlators $D^E$
 and $D^E_1$, given in \cite{15,17}

 \be D^E_1 (x) = \frac{6\alpha_s M_1 \sigma_f}{x} e^{-M_1x}
 \equiv\frac{A_1e^{-M_1 x}}{x} ;~~ x=\sqrt{\lambda^2+\nu^2},\label{A4}\ee with
 $$ \sigma_f =0.18~{\rm GeV}^2, ~~ M_1 = 1.4 ~{\rm GeV},$$

 \be D^E(x) = \frac{g^4 (N^2_c-1)}{2}0.108 \sigma^2_f  e^{-M_2x}
 ,\label{A5}\ee
 with $M_2=1.5$ GeV is the mass of the two-gluon gluelump with the account of
 perturbative interaction. As a result of integration in (\ref{A3}) of the
 forms (\ref{A4}) and (\ref{A5}) one obtains $\bar V_{\rm sat} (\infty)$ in the
 confinement phase
 \be \bar V_{\rm sat} (\infty) = \frac{A_1}{M_1^2} - \frac{4A_2}{M^3_2} =
 (0.432-0.415) ~{\rm GeV}~ \cong 17~{\rm MeV},\label{A^}\ee
 for $\alpha_s =0.4$. One can find $\bar V_{\rm sat}(r)$ for finite $r$   in
 the range $O(10$ MeV).

    \end{document}